# Gold on WSe$_2$ Single Crystal Film as a Substrate for Surface Enhanced Raman Scattering (SERS) Sensing

**Bablu Mukherjee,** [1,2,†] **Wei Sun Leong,** [1,†] **Yida Li,** [1] **Hao Gong,** [3] **Linfeng Sun,** [4] **Ze Xiang Shen,** [4] **Ergun Simsek,** [2] **and John T. L. Thong,** [1,*]

[1]Department of Electrical and Computer Engineering, National University of Singapore, Singapore 117583.
[2]Electrical and Computer Engineering, School of Engineering and Applied Science, The George Washington University (GWU), Washington, DC 20052, USA.
[3]Department of Materials Science & Engineering, National University of Singapore, Singapore 117575.

[4]Centre for Disruptive Photonic Technologies, School of Physical and Mathematical Sciences, Nanyang Technological University, Singapore 637371, Singapore

*Corresponding author's e-mail: elettl@nus.edu.sg

[†]These authors contributed equally to this work.

**Abstract:**

Synthesis and characterization of high-quality single-crystal WSe$_2$ films on highly-insulating substrate is presented. We demonstrate for the first time that the presence of gold nanoparticles in the basal plane of a WSe$_2$ film can enhance its Raman scattering intensity. The experimentally observed enhancement ratio in the Raman signal correlates well with the simulated electric field intensity using a three-dimensional electromagnetic software and theoretical calculation. This work provides guidelines for the use of two-dimensional WSe$_2$ films as a SERS substrate.

In the recent years, transitional metal dichalcogenide (TMD) materials have attracted tremendous attention due to their potential applications in valleytronics, flexible and low power electronics, optoelectronics and sensing devices **[1-6]**. One of the TMD materials, tungsten diselenide ($WSe_2$) has a structure of Se-W-Se covalently bonded in a hexagonal quasi-2D network configuration that is stacked by weak Van der Waals forces. It is a group VI TMD that exhibits a trigonal prismatic structure with an indirect band gap of 1.21 eV in the bulk form and which increases to 1.25 eV (direct) at monolayer thickness **[7]**. Besides having excellent electrical properties, $WSe_2$ shows great promise for optical sensing applications **[2-4, 8]**. Due to its exceptional optical properties, $WSe_2$ yields strong Raman signal compared to other TMD materials. Nevertheless, no work has reported that explores the potential of enhancing its Raman signal to render it a suitable surface-enhanced Raman spectroscopy (SERS) substrate. In view of SERS being a topic of growing research interest, substrate materials with strong Raman signal are being sought for sensing applications in analytical, biological and surface sciences.

In this work, we first synthesize large triangular/hexagonal $WSe_2$ single crystal films on c-face sapphire substrates *via* chemical vapor deposition (CVD) techniques. We demonstrate that the Raman signal of $WSe_2$ films can be enhanced easily by Au decoration making it a competitive SERS substrate for sensing device applications. An electromagnetic wave simulation and theoretical calculations were performed to quantitatively explain the experimentally observed enhancement in Raman scattering of our $WSe_2$ film arising from the Au decoration.

**Results and Discussion**

We have been able to synthesize **[5,6]** single crystal $WSe_2$ film continuous over a length of more than 170 μm. Figure 1a illustrates the experimental setup that was used to grow the $WSe_2$ films on c-face sapphire substrate via a CVD approach (see Experimental Methods for details). Figure 1b shows a photograph of the test tube containing 0.3 grams of $WSe_2$ powder at its closed end while the sapphire substrate was placed about 4 cm away from the powder source. It is worth noting that the growth of $WSe_2$ films on c-face sapphire substrate is highly dependent on the substrate position and temperature.

Apart from this, we also attempted to grow WSe$_2$ films on both a- and r-face sapphire substrates, but there was no sign of growth on both types of the sapphire substrates. Figure 1c and the inset of Figure 1d show the optical images of our synthesized WSe$_2$ films, which appear reasonably uniform and exhibit either triangular or hexagonal shapes with a lateral dimension of 100 to 200 μm. In order to check the uniformity of the synthesized WSe$_2$ films, we performed a large area Raman scan using a WITec alpha 300R system with a 532 nm (2.33 eV) laser excitation source and a step size of 160 nm. The laser power at the sample was kept below 0.1 mW which did not give rise to noticeable sample heating. As can be seen from the inset of Figure 1d, the intensity map of the E$^1_{2g}$ band is uniform throughout the large triangular WSe$_2$ film with a lateral size of more than 170 μm. In short, we have synthesized large triangular/hexagonal WSe$_2$ films on c-face sapphire substrates that are of uniform thickness.

In order to examine the number of layers of our CVD-grown WSe$_2$ films, we further characterized them through Raman analysis and atomic force microscopy. It has been reported that monolayer WSe$_2$ nanosheets on sapphire substrate show A$_{1g}$ mode vibration at 264 cm$^{-1}$ and two dominant peaks are observed around 250 cm$^{-1}$ in various samples from monolayer to bulk on SiO$_2$ (300 nm)/Si substrate **[5,9]**, while bulk WSe$_2$ exhibits two distinct Raman characteristic signals at 248 and 250.8 cm$^{-1}$ theoretically as well as experimentally **[10,11]**. Our as-synthesized WSe$_2$ film on sapphire substrate has two characteristic Raman peaks located at ~252 cm$^{-1}$ and ~260 cm$^{-1}$, which can be assigned to the in-plane vibrational E$^1_{2g}$ mode and the out-of-plane vibrational A$_{1g}$ mode, respectively (Figure 1d). In addition, another prominent Raman band was detected at ~309 cm$^{-1}$ for our as-synthesized WSe$_2$ film, which is an indication of interlayer interaction for layered 2D material **[6]**. Furthermore, the thickness of our WSe$_2$ film is measured to be 1.6 nm (Figure S1), and hence we believe that our CVD-grown WSe$_2$ film is a bilayer.

We also observed crystallinity of our CVD-grown WSe$_2$ films in a transmission electron microscope (TEM). The TEM image in Figure 2a shows the periodic atom arrangement of the WSe$_2$ film, indicating that the WSe$_2$ film is highly crystalline. Figure 2b shows the selected area electron diffraction (SAED) pattern taken on the WSe$_2$ film with an aperture size of ~200 nm. The high-resolution TEM image and its corresponding SAED pattern indicate hexagonal lattice structure with a spacing of 0.38 and 0.34 nm that can be assigned to the (100) and (110) planes, respectively. Apart from that, the chemical composition of the synthesized film is determined to be tungsten (W) and selenium (Se) with an atomic ratio of 1:2 by using an energy-dispersive X-ray spectroscopy (EDX) detector that is attached to the TEM (Figure 2c). The carbon and copper peaks in EDX spectrum can be attributed to the thin carbon film and copper mesh of TEM grid holder, respectively. These findings confirmed that the synthesized film is a single crystal WSe$_2$ film.

Thus far, we have confirmed that our CVD-grown WSe$_2$ film on c-face sapphire substrate is a bilayer and single crystal, and more importantly, it shows prominent Raman characteristic peaks. In order to transform the synthesized WSe$_2$ film into a promising SERS substrate, we decorated it with gold particles *via* 2 simple steps: Au film deposition and annealing. A thin layer of Au film (~5 nm (Figure S2)) was first deposited on the CVD-grown WSe2 film, followed by annealing in an inert environment at 550 $^o$C for 3 h to form Au nanoparticles (NPs) on the WSe$_2$ films (see Methods for details). Figure S3(a) shows the SEM image of a WSe$_2$ film on sapphire substrate with covered by Au NPs. Higher magnification SEM image (Figure S3a) and the NPs distribution graph (Figure S3b) estimate a clearer view of Au NPs with average size of ~ 9-12 nm in diameter, and the average separation between two particles ~ 13-20 nm, which were formed on the WSe$_2$ film. Figure 3a shows a high resolution Raman intensity map of the $E^1_{2g}$ band of a typical WSe$_2$ film that was half-decorated with Au NPs. It is obvious that the Raman intensity of the Au-decorated WSe$_2$ film is higher compared to that of bare WSe$_2$ film. Figure 3b shows the Raman spectrum of both Au-decorated and bare WSe$_2$ films and its inset shows their relative intensity difference in $E^1_{2g}/A_{1g}$. Basically, the main characteristic Raman peak intensity of the WSe$_2$ film has been enhanced by ~ 1.14 times by Au NPs under the excitation wavelength of 532 nm. The $E^1_{2g}/A_{1g}$ peak intensity ratio is increased from 1.15 to 1.2 after Au NPs coating on the WSe$_2$ film. On the other hand, Figure 3c shows the optical image of a typical hexagonal WSe$_2$ film fully decorated with Au NPs, while Figure 3d shows its Raman intensity map of the $E^1_{2g}$ band. It can be clearly seen that the Raman intensity is uniform throughout the Au-decorated WSe$_2$ film. Figure S5(a) and (b) show the Lorentzian fitting of Raman spectra obtained from bare WSe$_2$ film and Au NPs coated WSe$_2$ film on sapphire substrate, respectively. From Lorentzian fitting of the peaks and their peaks position, it can be observed that the separation between the peaks A$_{1g}$ and E$^1_{2g}$ is increased ~ 0.2 cm$^{-1}$ for Au NPs coated WSe$_2$ film sample compared with pristine WSe$_2$ sample, which basically falls in instrumental lowest detectable range. However it can be noted that both peaks' position are shifted towards lower frequency number (blue shift) for Au NPs coated WSe$_2$ sample compared with pristine WSe$_2$ sample, which is due to *p*-doping caused by electron transfer from the WSe$_2$ film to Au NPs [14-16]. In brief, Au-decoration enhances Raman signal of WSe$_2$ film making it a more promising candidate for SERS sensing at excitation wavelength of 532 nm.

We have carried out Raman scan on WSe$_2$ film with and without Au NPs coating for two other excitation wavelengths of 488 and 633 nm using WITec alpha 200R confocal Raman microscope as shown in Figure S6 (a) and (b), respectively. However, the prominent main characteristic Raman peak intensity ratios (|I$_{Au+WSe2}$/I$_{WSe2}$|) with and without Au NP coating on WSe$_2$ film are 1.01 and 1.04 for excitation wavelength of 488 nm and 633 nm, respectively. In order to understand the wavelength dependent nature of SERS capability of Au-decorated WSe$_2$ films and to investigate how electric field distribution is changed by the presence of Au NPs, we performed a set of simulations and calculations as follows.

When excited with an incident electromagnetic wave, where the associated electric field intensity is **E** = **n**$_e E_0$ cos($\omega t$), the dipole moment induced on an atom (or a nanoparticle) is given by **P** = $\alpha$ **E**, where $\alpha$ is the polarizability of the atom, $\omega$ is the angular frequency, and **n**$_e$ is a unit vector representing the polarization of electric field. Under such excitation, atoms vibrate around their equilibrium position and the physical displacement can be approximated as $dQ = Q_0$cos($\omega_v t$), where $Q_0$ is the maximum displacement and $\omega_v$ is the vibration frequency. Since the displacements for these planarly confined atoms are very short, their polarizability can be well approximated with a Taylor serious expansion, such that $\alpha = \alpha_0 + \frac{\partial \alpha}{\partial Q} dQ$, where $\alpha_0$ is the polarizability of the atom at the equilibrium position. Then the induced dipole moment intensity can be written as

$$P = \alpha_0 E_0 \cos(\omega t) + \frac{\partial \alpha}{\partial Q} \frac{Q_0 E_0}{2} \{\cos((\omega - \omega_v)t) + \cos((\omega + \omega_v)t)\} \qquad (1)$$

Basically, this equation means that all Rayleigh (first term), Stoke (second term), and anti-Stoke (third term) scattering components are proportional to the incident field strength. The experimentally observed Raman peak intensity enhancement informs us that when the $WSe_2$ layer is decorated with Au nanoparticles, the electric field intensity inside the $WSe_2$ layer changes and this change is in the positive direction (increase) for the wavelength range of our interest.

The field distribution throughout the $WSe_2$ coated sapphire substrates can be calculated analytically. For the Au nanoparticle decorated samples, one can either use the well-known coupled-dipole approximation (CDA) **[17]** or a commercially available electromagnetic software package **[18]**. In this work, we followed both approaches assuming Au cylindrical nanoparticles (with a diameter of 9 nm and a height of 6 nm) are periodically aligned on top of a bilayer $WSe_2$ coated sapphire substrate. The inter-particle spacing is 12 nm along the *x*- and *y*-axes. The frequency dependent complex permittivity of $WSe_2$ is taken from *Reference 19*, the refractive index of sapphire is assumed to be 1.768. For the optical constants of gold, experimental values are used rather than the Drude model to eliminate any concern regarding the selection of appropriate values for plasmon and relaxation frequencies **[20,21]**.

For the CDA approach, we first calculate the average electric field intensity, $E_{ave} = \sum_N (E_{LM}^{inc} + E_{LM}^{ind} - E_{LM}^{scat})/N$, on the $WSe_2$ surface (at -6 < X < 6 nm, -6 < Y < 6 nm), where $E_{LM}^{inc}$ is the layered medium incident electric field, $E_{LM}^{ind}$ is the electric field created by the induced dipoles, and $E_{LM}^{scat}$ is the back-scattered electric field. It should be noted that both for $E_{LM}^{ind}$ and $E_{LM}^{scat}$, the layered medium polarizability factors are used, as explained in *Reference 17*. Then, the ratio of $E_{ave}$ to average layered medium electric field intensity, $E_{ave}^{inc} = \sum_N (E_{LM}^{inc})/N$, gives the approximate field enhancement ratio. The black line in Figure 4 shows this ratio as a function of wavelength for $400 \leq \lambda \leq 700$ nm. This theoretical result suggests that the field is enhanced when the wavelength is 475 nm or higher and the maximum possible enhancement ratio is ~1.21 at the wavelength of ~540 nm.

For the numerical approach, we use Waveneology **[18]** to calculate the field enhancement ratio by utilizing mesh setting of 200 points per wavelength. Since Wavenology is a full wave electromagnetic solver, it is expected to provide more realistic results, as it makes no approximation for the polarizability of the nanoparticles. We calculate the electric field distributions with and without nanoparticles for the same physical parameters utilized in the CDA solution by assigning periodic boundary conditions at the ± X and ± Y boundaries and perfectly matched layers at the ± Z boundaries. Figure 5 shows the magnitudes of electric field distributions at three wavelength values (488, 532, and 633 and) over the $WSe_2$ surface with and without Au nanoparticles. The ratios of former average field intensities to latter ones yield the enhancement ratios of 1.014, 1.132, and 1.029. These numbers, marked with blue circular patch on Figure 4, shows a good agreement with the experimental results of prominent Raman peak intensity ratio, depicted with red stars in the same Figure 4.

In summary, large $WSe_2$ films were synthesized on highly-insulating sapphire substrate using CVD technique. We carefully examine the synthesized films using Raman spectroscopy, TEM, EDX and AFM and confirmed that our CVD-grown $WSe_2$ films on c-face sapphire substrate are single crystal bilayer $WSe_2$ films with prominent Raman characteristic peaks. We demonstrate both experimentally and numerically that Raman signature of the $WSe_2$ films can be enhanced by Au decoration due to surface plasmon resonance. To sum up, gold on $WSe_2$ single crystal film holds promise as a SERS substrate which could be suitable for sensing application.

**Experimental Methods**

*Growth and Characterization of $WSe_2$ Films*

In our experiments, the synthetic route using chemical vapor deposition (CVD) technique is described as follows. Pure $WSe_2$ powder (Sigma Aldrich, purity 99.8 %) was used as source materials. C-face sapphire substrates were placed inside the one-end open quartz-glass tube, where small amount (~ 0.3 grams) of WSe$_2$ source powder was loaded at the closed end of the tube as shown in the picture (Figure 1b). The tube was inserted in a horizontal quartz tube placed in a conventional tube furnace such that the substrate was set at lower temperature region of the source powders and the distance between them was about 4 cm. Then the quartz tube was evacuated to a base pressure ~ $10^{-3}$ mbar for 2 h by a vacuum pump and subsequently was filled with the mixture of argon (Ar) with 5% H$_2$ gas. The gas was allowed to flow for 1 hr after flushing the tube 2-3 times. After that, the furnace was heated under the mixture gas of Ar and 5% H$_2$ at a flow rate of 100 sccm (standard cubic centimeters per minute). When the temperature reached 950 $^{o}$C (heating rate: 30 $^{o}$C min$^{-1}$), the pressure of Ar carrier gas was maintained at ~ 2 mbar during synthesis for 15 mins. After the reaction was terminated, the substrates were taken out when the temperature of the furnace cooled down to room temperature.

The morphology, structure and chemical composition of the as-synthesized nanostructures were characterized using atomic force microscopy (AFM, Vecco D3000 NS49 system), transmission electron microscopy (TEM, JEOL, JEM-2010F, 200 kV), energy-dispersive X-ray spectroscopy (EDX) equipped in the TEM and Raman spectroscopy. All Raman analysis in this study were performed using a Raman system (WITec alpha 300R) with a 532 nm laser excitation source and laser spot size of ~320 nm (x100 objective lens with numerical aperture 0.9). The laser power at the sample was kept below 0.1 mW which did not give rise to noticeable sample heating [12]. All Raman mappings were conducted with a step size of 160 nm. The spectral resolution was ≤ 1.5 cm$^{-1}$ (using a grating with 1,800 grooves mm$^{-1}$) and each spectrum was an average of 10 acquisitions (0.1 s of accumulation time per acquisition).

*Formation of Au NPs on the CVD-grown WSe$_2$ Films*

A metal shadow mask was used such that some of the CVD-grown WSe$_2$ films on c-face sapphire substrates were partially exposed. Subsequently, Au film was directly deposited on the sample via thermal evaporation at a rate of 0.05 nm/s for 3-5s at a chamber base pressure of 3 x 10$^{-6}$ mbar, and the thickness of Au film was measured to be ~5nm by AFM (Figure S2). The sample was then annealed at 550 °C for 3 h in an Ar gas environment to allow formation of Au NPs on the CVD-grown WSe$_2$ films [13]. As can be seen in Figure S3, the size of each Au NP is ~ 9-12 nm in diameter with a spacing of 13-20 nm between NPs.

**References**

1. Lin J, Li H, Zhang H, Chen W 2013 Plasmonic Enhancement of Photocurrent in MoS$_2$ Field-Effect-Transistor *Appl. Phys. Lett.* **102** 203109

2. Ross JS, *et al.* 2014 Electrically Tunable Excitonic Light-Emitting Diodes Based On Monolayer WSe$_2$ P-N Junctions *Nat. Nanotechnol* **9** 268-272

3. Das S, Appenzeller J 2013 WSe$_2$ Field Effect Transistors with Enhanced Ambipolar Characteristics *Appl. Phys. Lett.* **103** 103501

4. Zhao W, *et al.* 2012 Evolution of Electronic Structure in Atomically Thin Sheets of WS$_2$ and WSe$_2$ *ACS Nano* **7** 791-797

5. Kai X, Zhenxing W, Xiaolei D, Muhammad S, Chao J, Jun H 2013 Atomic-Layer Triangular WSe$_2$ Sheets: Synthesis and Layer-Dependent Photoluminescence Property *Nanotechnology* **24** 465705

6. Huang J-K, *et al.* 2013 Large-Area Synthesis of Highly Crystalline WSe$_2$ Monolayers and Device Applications *ACS Nano* **8** 923-930

7. Sahin H, *et al.* 2013 Anomalous Raman Spectra and Thickness Dependent Electronic properties of WSe$_2$ *Phys. Rev. B* **87** 165409

8. Jones AM, *et al.* 2013 Optical Generation of Excitonic Valley Coherence in Monolayer WSe$_2$ *Nat. Nanotechnol* **8** 634-638

9. Zeng H, *et al.* 2013 Optical signature of symmetry variations and spin-valley coupling in atomically thin tungsten dichalcogenides *Sci. Rep.* **3** 1608 1-5

10. Tonndorf P *et al.* 2013 Photoluminescence emission and Raman response of monolayer MoS$_2$, MoSe$_2$, and WSe$_2$ *Opt. Express* **21** 4908-4916

11. Ding Y, Wang Y, Ni J, Shi L, Shi S, Tang W 2011 First principles study of structural, vibrational and electronic properties of graphene-like $MX_2$ (M=Mo, Nb, W, Ta; X=S, Se, Te) monolayers. *Physica B: Condensed Matter* **406** 2254-2260

12. Leong WS, Nai CT, Thong JTL 2014 What Does Annealing Do to Metal–Graphene Contacts? *Nano Lett.* **14** 3840-3847

13. Bowker M.*, et al.* 2013 Encapsulation of Au Nanoparticles on a Silicon Wafer During Thermal Oxidation *J Phys. Chem. C* **117** 21577-21582

14. Yumeng Shi *et al.*, 2013 Selective Decoration of Au Nanoparticles on Monolayer $MoS_2$ Single Crystals *Sci. Rep.* **3** 1839

15. Shao Su *et al.* 2014 Creating SERS Hot Spots on $MoS_2$ Nanosheets with in Situ Grown Gold Nanoparticles *ACS Appl. Mater. Interfaces* **6(21)** 18735–18741

16. Chang-Hsiao Chen *et al.* 2014 Hole mobility enhancement and p-doping in monolayer $WSe_2$ by gold decoration *2D Materials* **1** 034001

17. Simsek E 2010 Full analytical model for obtaining surface plasmon resonance modes of metal nanoparticle structures embedded in layered media *Optics Exp.* **18(2)** 1722-1733

18. Wavenology from Wave Computation Technologies Inc. Durham NC

19. Li S-L, Miyazaki H, Song H, Kuramochi H, Nakaharai S, Tsukagoshi K 2012 Quantitative Raman Spectrum and Reliable Thickness Identification for Atomic Layers on Insulating Substrates *ACS Nano* **6** 7381-7388

20. Simsek E 2013 Improving Tuning Range and Sensitivity of Localized SPR Sensors With Graphene *IEEE Phot. Tech. Lett.* **25(9)** 867-870

21. Raki AD, Djurisic AB, Elazar JM, Majewski ML 1998 Optical properties of metallic films for vertical-cavity optoelectronic devices *Appl. Opt.* **37** 5271-5283
10

**Acknowledgements**

This project is supported by grant R-263-000-A76-750 from the Faculty of Engineering, NUS, and grant NRF2011NRF-CRP002-050 from the National Research Foundation, Singapore.

**Additional information**

Supplementary Information: available.

Competing financial interests: The authors declare no competing financial interests.

**FIGURES**

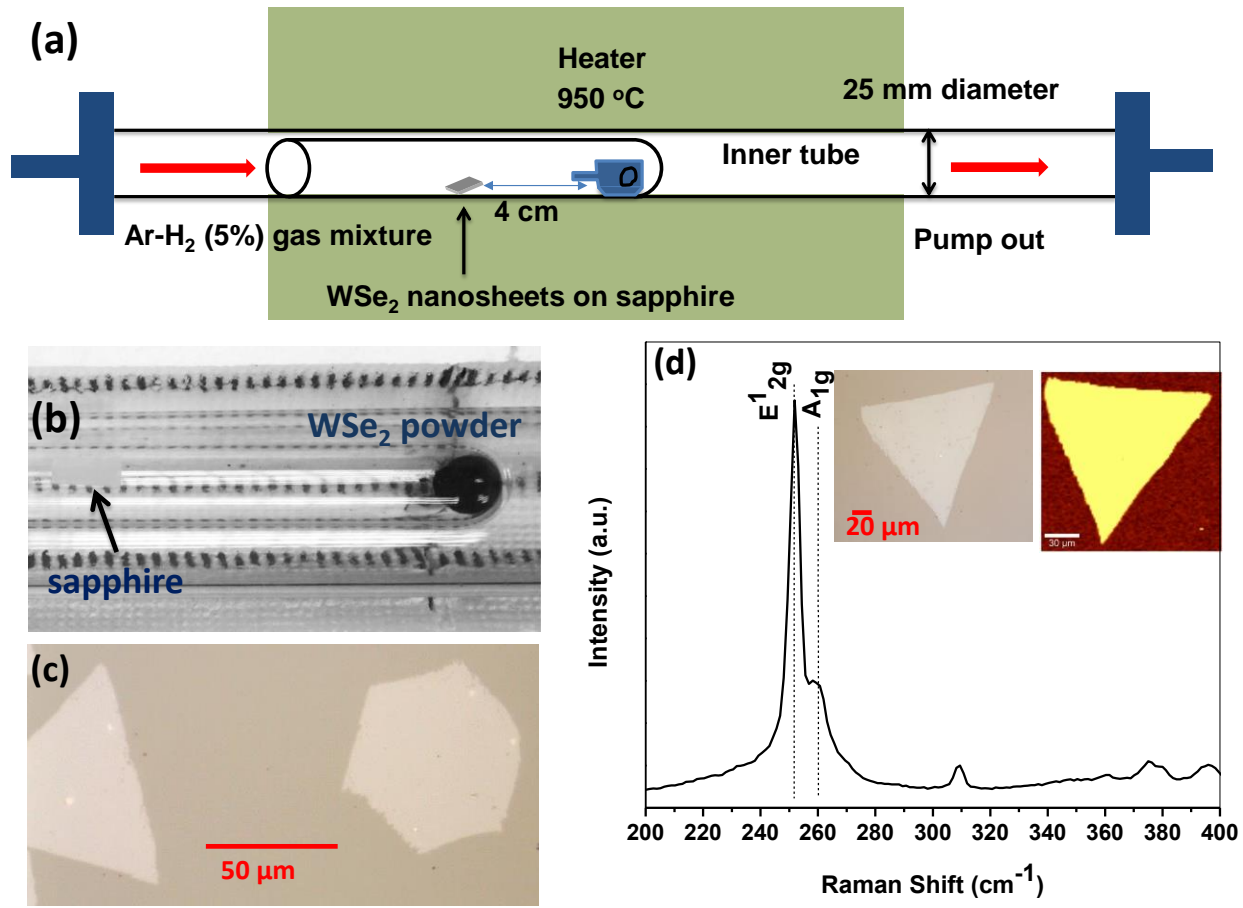

**Figure 1.** (a) Schematic representation of the experimental setup that we used to grow the WSe$_2$ films on c-face sapphire substrate. (b) Photo of the test tube containing 0.3 grams of WSe$_2$ powder at its closed end and the sapphire substrate was placed at about 4 cm away from the powder source. (c) Optical image of the typical as-synthesized WSe$_2$ films on a c-face sapphire substrate. (d) Raman spectrum of the typical WSe$_2$ film in its insets showing optical image of an individual WSe$_2$ triangular film (left) and the corresponding Raman intensity map of the E$^1_{2g}$ band (right).

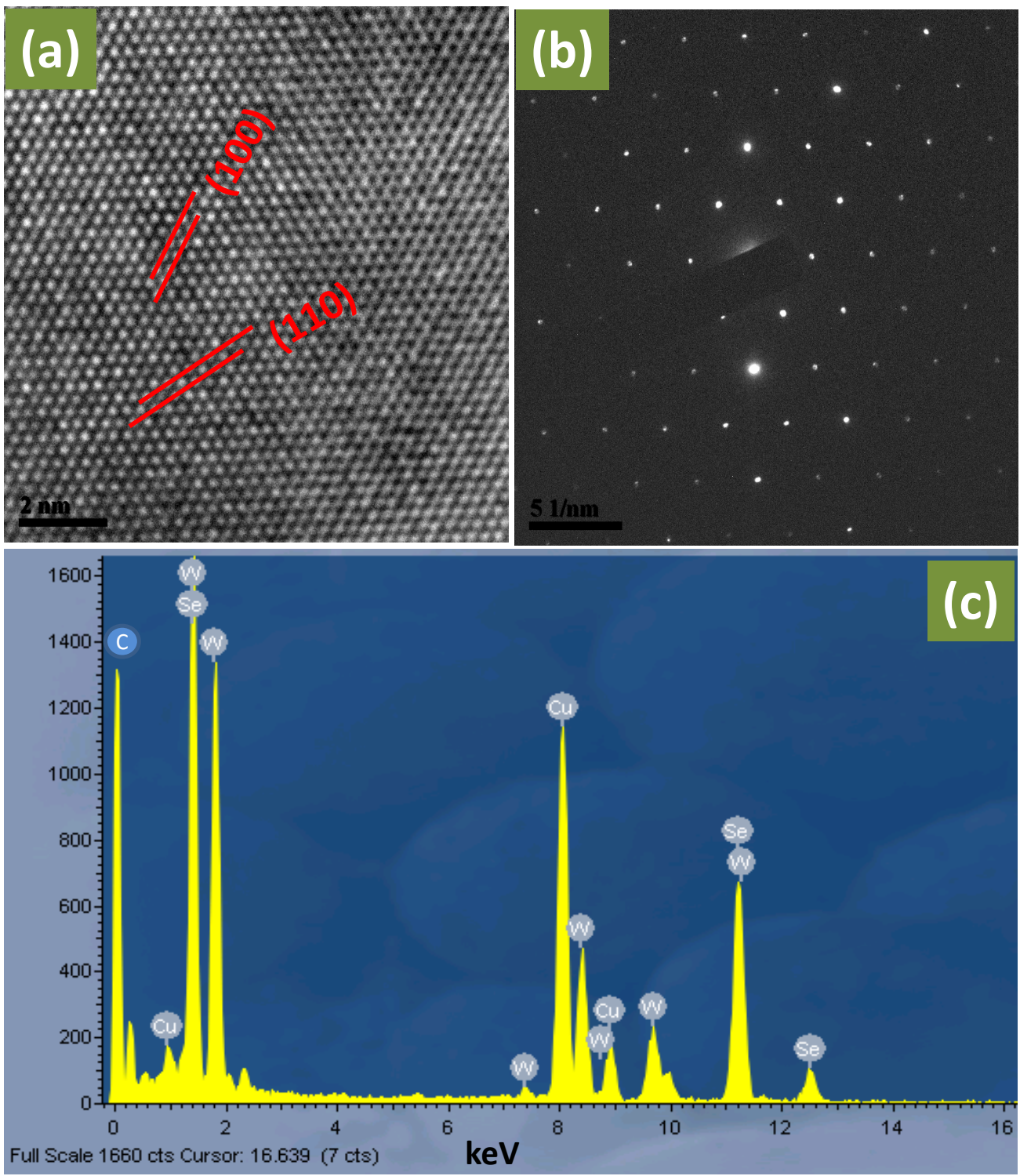

**Figure 2.** (a) High-resolution TEM image of a typical CVD-grown WSe$_2$ film. (b) The corresponding selected area electron diffraction (SAED) pattern. (c) Energy-dispersive X-ray (EDX) spectrum of the CVD-grown WSe$_2$ film.

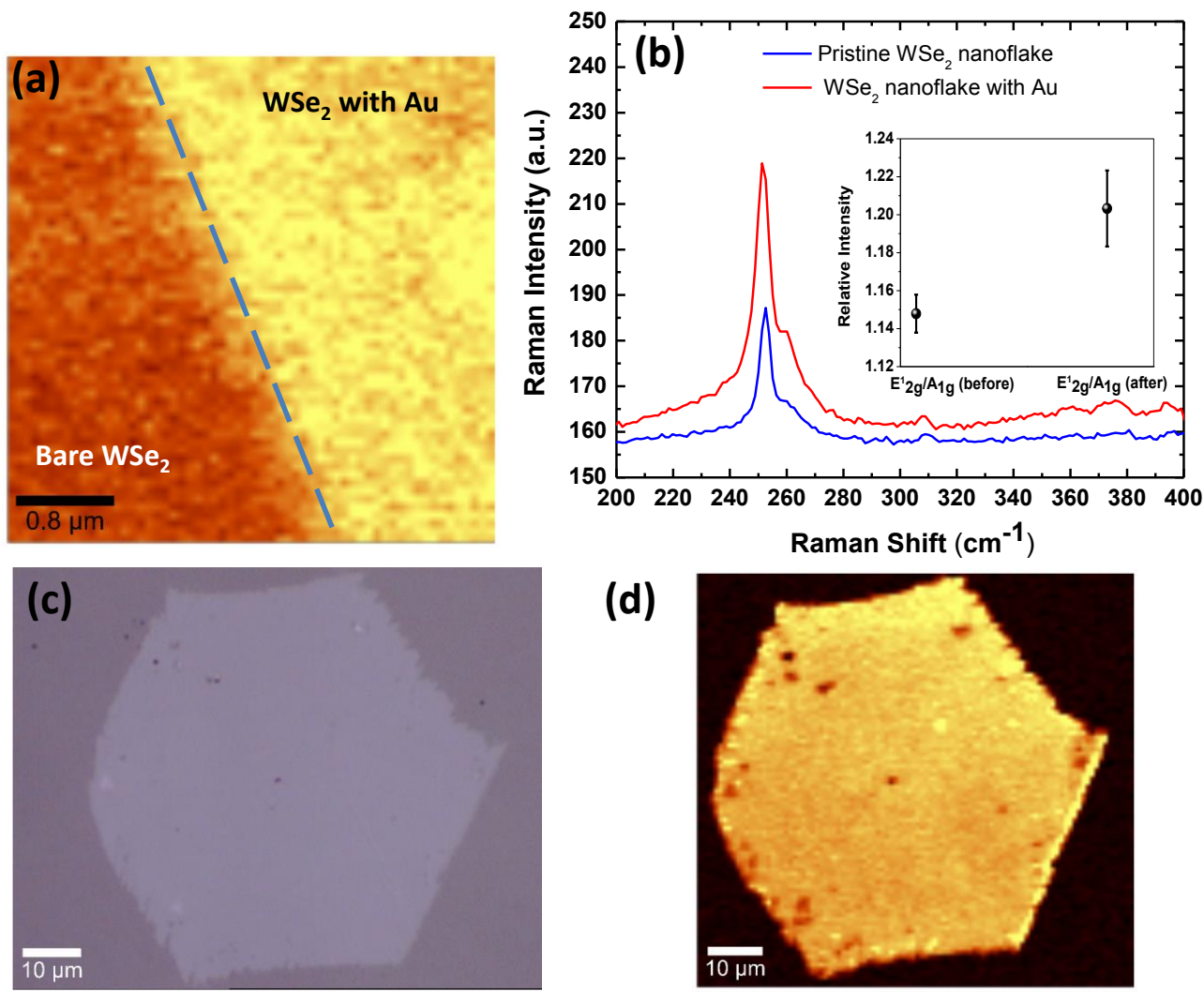

**Figure 3.** (a) High resolution Raman intensity map of the $E^1_{2g}$ band of a typical WSe$_2$ film that was half-decorated by Au NPs. (b) Raman spectra of both Au-decorated and bare WSe$_2$ films and its inset shows their relative intensity difference in $E^1_{2g}/A_{1g}$ at excitation wavelength of 532 nm. (c) Optical image of a typical hexagonal WSe$_2$ film fully decorated by Au NPs, and (d) its Raman intensity map of the $E^1_{2g}$ band.

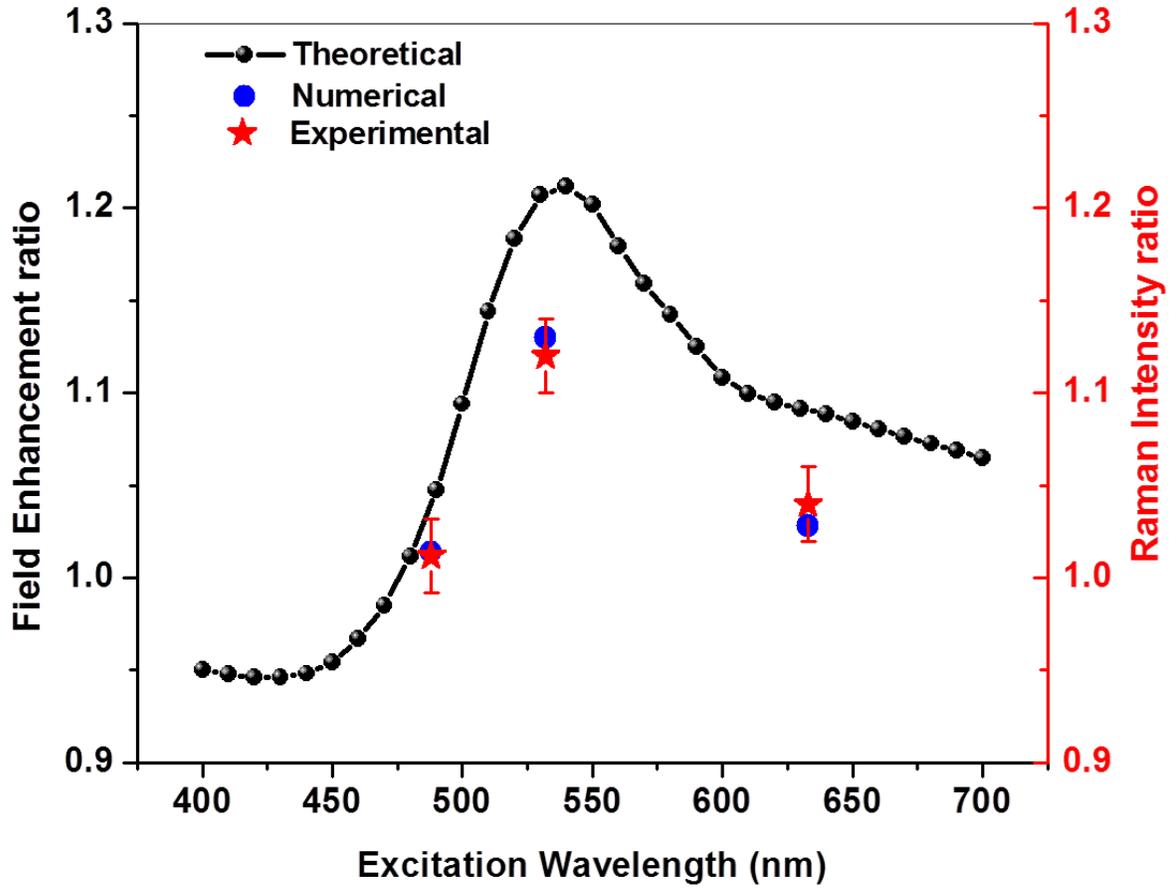

**Figure 4**. Theoretically and numerically calculated field enhancement ratio *versus* excitation wavelength, where the Raman intensity ratio of the structure consist with and without Au NPs on top of WSe$_2$ film/Sapphire substrate is also plotted in same wavelength range.

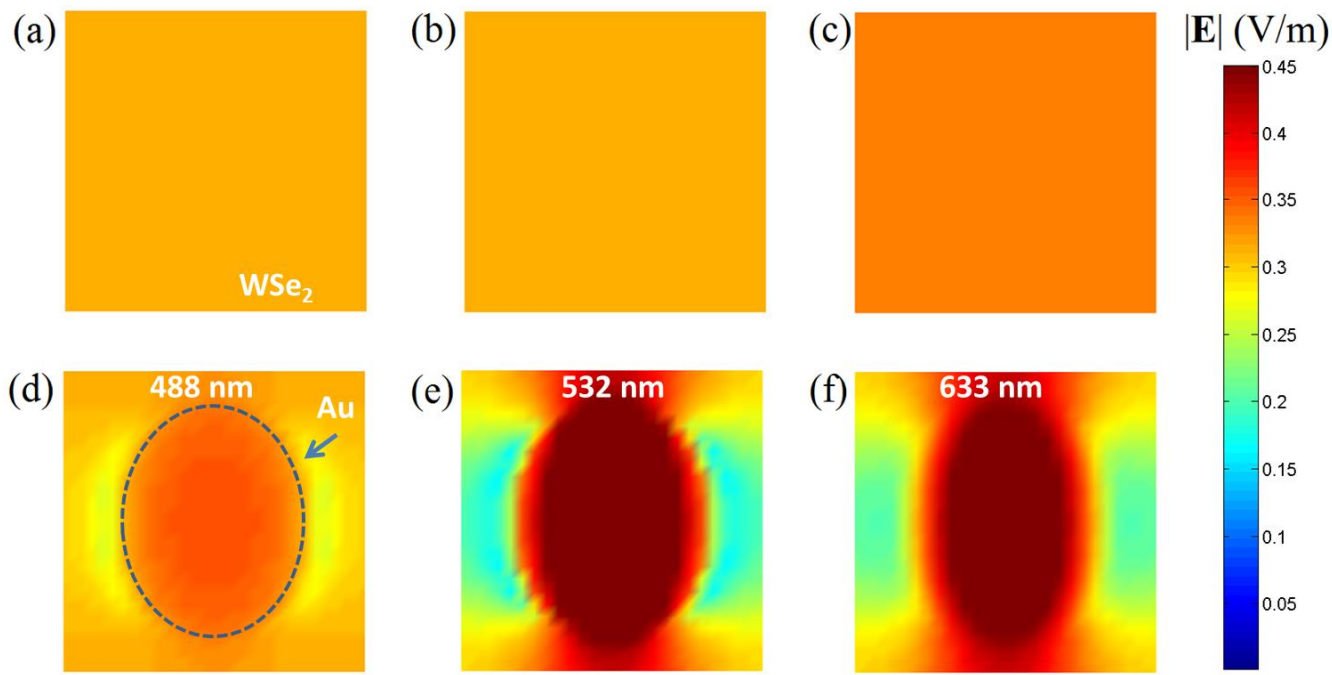

**Figure 5.** (a, b, c) Simulated magnitude of the electric field (|E|) distribution on WSe$_2$ plane under excitation wavelength of 488nm, 532nm and 633nm, respectively. (d, e, f) |E| distribution in the plane of interface between Au NP and WSe$_2$ film corresponding those excitation wavelength as labeled.